# Band-selective Holstein polaron in Luttinger liquid material $A_{0.3}$MoO$_3$ ($A$ = K, Rb)

L. Kang[1,*], X. Du[1,*], J. S. Zhou[1], X. Gu[1], Y. J. Chen[1], R. Z. Xu[1], Q. Q. Zhang[1], S. C. Sun[1], Z. X. Yin[1], Y. W. Li[2,3], D. Pei[4], J. Zhang[2], R. K. Gu[4], Z. G. Wang[5], Z. K. Liu[2,3], R. Xiong[6], J. Shi[6], Y. Zhang[5], Y. L. Chen[1,2,3,4,‡] and L. X. Yang[1,7,‡]

[1]*State Key Laboratory of Low Dimensional Quantum Physics, Department of Physics, Tsinghua University, Beijing 100084, China.*
[2]*School of Physical Science and Technology, ShanghaiTech University and CAS-Shanghai Science Research Center, Shanghai 201210, China.*
[3]*ShanghaiTech Laboratory for Topological Physics, Shanghai 200031, China.*
[4]*Department of Physics, Clarendon Laboratory, University of Oxford, Parks Road, Oxford OX1 3PU, UK.*
[5]*International Center for Quantum Materials, School of Physics, Peking University, Beijing 100871, China.*
[6]*Department of Physics, Wuhan University, Wuhan 430072, China.*
[7]*Frontier Science Center for Quantum Information, Beijing 100084, China.*

[*]*These authors contribute equally to this work.*

[‡]*Email address: LXY: lxyang@tsinghua.edu.cn, YLC: yulin.chen@physics.ox.ac.uk*

**(Quasi-)one-dimensional systems exhibit various fascinating properties such as Luttinger liquid behaviour, Peierls transition, novel topological phases, and the accommodation of unique quasiparticles (e.g., spinon, holon, and soliton, etc.). Here we study molybdenum blue bronze $A_{0.3}$MoO$_3$ ($A$ = K, Rb), a canonical quasi-one-dimensional charge-density-wave material, using laser-based angle-resolved photoemission spectroscopy. Our experiment suggests that the normal phase of $A_{0.3}$MoO$_3$ is a prototypical Luttinger liquid, from which the charge-density-wave emerges with decreasing temperature. Prominently, we observe strong renormalizations of band dispersions, which is recognized as the spectral function of Holstein polaron derived from band-selective electron-phonon coupling in the system. We argue that the strong electron-phonon coupling plays a dominant role in electronic properties and the charge-density-wave transition in blue bronzes. Our results not only reconcile the long-standing heavy debates on the electronic properties of blue bronzes but also provide a rare platform to study novel composite quasiparticles in Luttinger liquid materials.**

# Introduction

Dimensional confinement can strongly influence the electronic properties of many-body systems. In drastic contrast to two- or three-dimensional systems, low-energy particle-hole pair in one-dimensional system can be excited only if its wave vector $|q| \approx 0$ or $|q| \approx 2k_F$ with $k_F$ being the Fermi momentum of electrons (Fig. 1a). The reduction of scattering phase-space and electronic screening gives rise to rich fascinating properties, such as power-law correlation function and spin-charge separated collective bosonic excitations, as properly described by Luttinger liquid (LL) theory[1,2]. On the other hand, quasi-one-dimensional (Q1D) metals with well nested Fermi surface (FS) are highly susceptible toward a charge-density-wave (CDW) state with periodic lattice distortion and an energy gap near $E_F$ (Fig. 1b)[3], in which electron-phonon coupling (EPC) usually plays a crucial role. Despite the intensive theoretical efforts[4-7], the LL-to-CDW transition is rarely studied experimentally[8-10], mainly due to the lack of a suitable materials platform. Moreover, with stronger EPC, electrons can be dressed by local lattice distortion, forming novel composite quasiparticles ─ the alleged Holstein polarons that can strongly renormalize the electronic structure of the system and induce energy gaps and "flat bands" at $-n\Omega_0$ ($n = 1, 2…$) with $\Omega_0$ being the energy of the strongly coupled phonon involved in the formation of the polaron (Fig. 1c). Up to date, while polarons have been widely observed and well understood in two- or three-dimensional materials[11-16], they are yet to be discovered in Q1D materials.

On account of the assumptions of LL theory, it is essential for a real material to be a highly-anisotropic metal with linear band dispersion. Several crystalline materials meeting these criteria have been demonstrated to be LL candidates, such as 1D cuprate[17], Lithium purple bronze[18], Chromium-based superconductor[19], and organic metallic systems[8, 20]. Among the various Q1D systems, molybdenum blue bronze $A_{0.3}MoO_3$ ($A$ = K, Rb, Tl) provides a suitable platform to investigate the LL-to-CDW transition. It exhibits a CDW transition at $T_{CDW} \approx 183$ K[21, 22] (Supplementary Fig.1), accompanied by a Kohn anomaly[23] and a sharp peak in the Lindhard function[24]. Angle-resolved photoemission spectroscopy (ARPES) measurements suggest an important role of FS nesting in the CDW transition[25, 26]. Nevertheless, the absence of the Fermi edge in the density of states[27, 28] has challenged this scenario and sparked heavy debates about non-Fermi liquid behavior [29-31] and (pseudo)gap above $T_{CDW}$ in the system[25, 27, 32, 33]. Moreover, the underlying microscopic interaction that is crucial for the CDW transition is also under heavy debate[25, 26, 33, 34]. These controversial results call for a comprehensive understanding of the single-particle spectral properties of the system.

In this work, we report a high-resolution laser-based ARPES study of the blue bronze $A_{0.3}MoO_3$ ($A$ = K and Rb). By tracking the band dispersion over a wide temperature range, we provide spectroscopic evidence of the CDW transition near $T_{CDW} \approx 183$ K. Above $T_{CDW}$, we observe a linear band dispersion with power-law scaling behavior of ARPES spectral weight, which can be precisely described by the LL model[35]. Below $T_{CDW}$, the LL property of the spectra revives after the CDW gap is suppressed by surface doping of Rubidium (Rb) atoms. Interestingly, we observe strong renormalizations of the band dispersions, which are identified as the spectral function of Holstein polaron derived from band-selective EPC in the system. We conclude that the strong EPC[22, 34, 36, 37], in addition to the FS nesting[25, 26], plays an important role in electronic properties and the LL-to-CDW transition in blue bronzes. Our results not only help understand the long-standing mysteries in blue bronzes, including the non-Fermi liquid behavior, the pseudogap above $T_{CDW}$, and the mechanism of the CDW transition, but also provide a rare platform to study the novel EPC in LL materials, which will also shed light on the understanding of rich physics in other Q1D materials.

## Results

**CDW transition in blue bronze.** As shown in Fig. 1d, $A_{0.3}MoO_3$ crystallizes into the monoclinic structure with space group $C2/m$. It consists of 1D chains of $MoO_6$ octahedrons extending along the [010] direction and $MoO_3$ layers intercalated by $A$ atoms stacking along the [2 0 -1] direction[38]. Figure 1e shows the Q1D FSs running along $b^*$ ($k_x$) measured at 80 K, consistent with the Q1D crystal structure and in good agreement with our calculation (Supplementary Fig. 3). Figures 1f and 1g show band dispersions along the $\Gamma$-$Y$ and $X$-$Z$ directions. The bonding ($B$) band shows a linear dispersion in a large energy range, while the anti-bonding ($AB$) band shows a relatively narrow band width with band bottom near -370 meV below $E_F$, in good agreement with our calculation (Supplementary Fig. 3).

To investigate the CDW transition in $A_{0.3}MoO_3$, we track the temperature evolution of the band dispersion using laser-based ARPES with superb resolutions, as shown in Fig. 2. At 304 K, we observe a strong suppression of the spectral weight near $E_F$ and the absence of Fermi edge. With decreasing temperature, the spectral weight is further suppressed and an energy gap ($\Delta_{CDW}$) gradually opens below $T_{CDW}$, as also shown by the energy distribution curves (EDCs) near the Fermi momenta $k_F$ of the $B$ and $AB$ bands (Figs. 2b and 2c). The CDW gap is about 42 meV at

83 K (Fig. 2d) and follows the BCS-type gap equation, in accordance with previous reports[27, 33, 39, 40] and the estimation based on our resistivity measurements (Supplementary Fig. 1).

**LL nature of the normal state.** As schematically shown in Fig. 3a, the spectral density in LLs shows a power-law behavior near $E_F$, which can be directly recognized by ARPES[18, 19, 31, 41, 42]. In blue bronzes, the absence of Fermi edge was previously attributed either to non-Fermi liquid behavior[30] or the pseudogap induced by the CDW fluctuations[25, 27, 32, 33]. However, compelling evidence for both scenarios is yet to be derived. In Fig. 3b, we approximate the EDCs integrated along Γ−Y near $E_F$ with a LL spectral function at finite temperatures[35]:

$$\rho(\epsilon, T) \propto T^\alpha \mathrm{Re}\left[(2i)^{\alpha+1} \mathrm{B}\left(\frac{\alpha + 1 + i\epsilon/\pi}{2}, -\alpha\right)\right], \quad (1)$$

where $\epsilon = (E - E_F)/k_B T$ is the scaled energy, B is the β function, and α is the LL anomalous exponent. The EDCs fit perfectly to the LL spectral function convoluted with the energy resolution in our experiment with α = 0.6 ± 0.1 (Supplementary Fig. 4) as shown in Fig. 3b. Moreover, equation (1) suggests a scaling relation of the EDCs, that is, $\rho(\epsilon, T)/T^\alpha$ is independent of the temperature, which is well established in Fig. 3c with the same α, compellingly proving the LL nature of blue bronzes in the normal state.

At temperatures slightly below $T_{\mathrm{CDW}}$, the spectra still fit nicely to the LL model after introducing an extra fitting parameter Δ accounting for the CDW gap in the spectral function (Supplementary Fig. 5), suggesting the intimate relationship between the CDW state and the LL phase. Figure 3d shows the temperature evolution of α. It increases slightly with decreasing temperature (above $T_{\mathrm{CDW}}$), in contrast to the two-band model proposed in $\mathrm{Li_{0.9}Mo_6O_{17}}$[18, 43]. Far below $T_{\mathrm{CDW}}$, the EDCs cannot be fitted to the LL model due to the strong modification of the

LL phase by the CDW ordering (Supplementary Fig. 5). Interestingly, the LL property revives after the CDW gap is suppressed by doping Rb atoms on the sample surface as shown by the shift of the leading edge of the EDCs in Fig. 3e and 3f, which saturates to a value of about 20 meV (Fig. 3f and Supplementary Fig. 6). After that, the spectrum again fits well to the LL model as shown in the inset of Fig. 3f, further proving our conclusion that the normal state of the blue bronze is a LL.

**Spectral function of Holstein polaron.** To understand the microscopic interactions underlying the LL-to-CDW transition, we perform further spectroscopic analyses in Fig. 4. Figure 4a shows the ARPES spectrum of the *AB* band (left) and its second derivative with respect to the energy (right). Interestingly, we observe a flat band near -150 meV and a discontinuity in the dispersion near -170 meV, which are clearly resolved in the EDCs (Fig. 4b), but absent in the *ab initio* calculation (Supplementary Note 3) and previous ARPES measurements. For more detailed inspection of the spectral function, we approximate the MDCs of the *AB* band with Lorentzians. Despite the strong suppression of the spectral weight by the LL nature of the system, we can still extract the band dispersion and the spectral broadening $\Delta k(E)$ near $E_\mathrm{F}$. As shown in Figs. 4c and 4d, we observe a kink-like structure near -85 ± 10 meV.

The doubled energy positions of the band renormalizations at ≈ -85 meV and ≈ -170 meV mimic the characteristic spectral function of Holstein polarons that are usually manifested by a series of band renormalization at $-n\Omega_0$ ($n$ = 1, 2,…), where $\Omega_0$ is the energy of the phonon mode involved in the formation of the polaron (Fig. 1c)[11, 13, 44]. To verify the formation of Holstein polaron in the blue bronze, we simulate its spectral function by a simple momentum-average model (see Methods for details)[11, 44]. Considering the energy position of the kink-like structure and the dispersion discontinuity, we adopt a phonon mode with energy of ≈ 85 meV in the simulation[45, 46]. Figures 4e-g elucidate the evolution of the band renormalizations with increasing EPC parameter λ. At λ ≈ 0.6 (Fig. 4g), both the flat band and the kink-like structure are excellently reproduced by our simple approximation (also see the EDCs of the simulated spectrum in Fig. 4h), confirming the formation of Holstein polaron in the system. It is noteworthy that the band renormalization near -85 meV is concealed by the suppression of the spectral weight near $E_F$ and is instead manifested by a kink-like structure in our experiment and simulation.

## Discussion

Although the electronic properties are radically different in Fermi liquids and LLs, we observe the formation of Holstein polaron in a LL material, similar to that in Fermi liquids[11]. To understand the mechanism of Holstein polaron in blue bronzes, it is crucial to reveal the nature of the phonon mode involved. The phonon mode at ≈ 85 meV is identified as the bridge Mo-O-Mo stretching mode as observed in Raman scattering and infrared spectroscopic experiments[45, 46]. Due to the Q1D crystal structure of blue bronzes, this phonon mode is spatially localized perpendicular to the atomic chains, thus is short-range in nature. Along the atomic chains, however, it can be relatively extended. The size or the coherence length of the Holstein polaron can be estimated by $l = \tau v_p$, where $\tau$ and $v_p$ are the life time and velocity of the quasiparticles. $\tau$ and $v_p$ can be obtained from the spectral broadening and Fermi velocity of the

*AB* band respectively. With $v_p \approx 3$ eV·Å and MDC width $\Delta k = 0.12$ Å$^{-1}$ near $E_F$, we estimate $\tau \approx 11$ fs and thus $l \approx 8.3$ Å, slightly larger than the lattice constant along the atomic chains.

Moreover, the strong band renormalizations and the formation of Holstein polaron are absent in the *B* band and the *AB* band is much more broadened than the *B* band, suggesting an interesting band-selective EPC. It is very likely that the phonon mode at ≈ 85 meV involves mainly the Mo II atoms whose orbitals contribute significantly to the *AB* band (Supplementary Note 3), giving rise to the band-selective formation of Holstein polarons. Even so, the *B* band is also influenced by the strong EPC in the system. By quantitative analyses, we show that the electron self-energy of the *B* band near $E_F$ strongly depends on the temperature (Supplementary Note 7), further confirming the strong EPC in blue bronzes[22, 34, 36, 37]. Likewise, the estimated coherence length of the Holstein polaron is in good comparison with the CDW period of about 9 Å and the phonon mode involved in the Holstein polaron is intimately related to the CDW transition[33, 45, 46], suggesting that the strong EPC and its band-selectivity may play an important role in the electronic properties and the LL-to-CDW transition in blue bronzes.

Finally, it is worth to note that we do not observe clear evidence of the spin-charge separation. On the one hand, the LL phase is only well established in blue bronzes at high temperatures where the strong thermal fluctuation challenges the observation of spin-charge separation. On the other hand, the dispersions of the separated spinon and chargon are expected to be distinguishable only in LL systems with α much smaller than 0.5[19, 31]. In addition, the strong EPC may smear the spin-charge separation in the system[47, 48].

In summary, with high-resolution laser-based ARPES measurements, we reveal clear spectroscopic evidence of the LL nature of blue bronzes in the normal state. We observe Holstein polarons that strongly renormalize the band dispersion. We therefore conclude that the strong EPC plays an important role in the spectral properties and CDW transition of blue bronzes. Our results suggest that the physics in blue bronze may be properly described by the 1D Holstein model, in which EPC is a vital parameter in the phase diagram[7]. Our work not only reveals a LL-to-CDW transition in blue bronzes but also provides a rare platform to study the novel Holstein-polaron-induced EPC in LL materials.

## Methods

### ARPES

The $A_{0.3}MoO_3$ single crystals were grown by an electrolytic reduction method[49]. High-resolution ARPES measurements were performed at beamline 13U of National Synchrotron Radiation Laboratory (NSRL), China, beamline 9A of Hiroshima Synchrotron Radiation Center (HSRC), Japan, and Peking University, China. Data were collected with Scienta R4000 (DA30) electron analyzers at NSRL and HSRC (Peking University). The overall energy and angle resolutions were set to 15 meV and 0.2°, respectively.

### Laser-based ARPES

Laser-based ARPES measurements were performed using DA30L analyzers and vacuum ultraviolet 7 eV lasers in Tsinghua University and ShanghaiTech University, China. The overall energy and angle resolutions were set to 6 meV and 0.2°, respectively. The samples were cleaved *in situ* and measured under ultra-high vacuum below $1.0 \times 10^{-10}$ mbar. Surface Rb doping was performed *in situ* at 80 K using a SAES alkali-metal source after well outgassing. The current was set to 5.6 A.

**First-principles calculation**

Electronic structure calculations were performed using density functional theory (DFT)[50] with projected augmented wave method as implemented in the QUANTUM ESPRESSO package[51, 52]. The exchange-correlation functional was approximated within the Perdew-Burke-Ernzerhof scheme[53]. Experimental structural parameters were relaxed with a force threshold of 0.01eV/Å. The cutoff energy for the plane-wave basis was set to 600 eV and the Monkhorst-Pack k-point mesh of 9×9×9 was used to get a self-consistent charge density. Fermi surface was calculated with a denser mesh of 16×16×16 and checked by the tight-binding based calculation supplied by the Wannier90 code[54].

**Simulation of the spectral function of Holstein polaron**

The spectral function calculation of Holstein polaron is performed within momentum average approximation[44], which is demonstrated to feature high efficiency and accuracy[11, 44]. The model has a simple form of an electron coupled with a dispersionless phonon mode, i.e., Holstein polaron. The Green's function of the Holstein model is represented as:

$$G(\mathbf{k}, \omega) = \frac{1}{\omega - \epsilon_\mathbf{k} - \Sigma_{MA}(\omega) + i\eta},$$

where $\omega$ is the energy, $\epsilon_\mathbf{k}$ is the bare dispersion, $\eta$ is the spectral broadening parameter, and the momentum-averaged self-energy $\Sigma_{MA}(\omega)$ has simple analytic form for 1D systems:

$$\Sigma_{MA}(\omega) = \cfrac{g^2 \Omega_0 \overline{g_0}(\omega - \Omega_0)}{1 - \cfrac{2g^2 \overline{g_0}(\omega - \Omega_0)\overline{g_0}(\omega - 2\Omega_0)}{1 - \cfrac{3g^2 \overline{g_0}(\omega - 2\Omega_0)\overline{g_0}(\omega - 3\Omega_0)}{1 - \cdots}}},$$

$$\overline{g_0}(\omega) = \frac{\text{sgn}(\omega)}{\sqrt{(\omega + i\eta)^2 - 4t^2}}$$

where $\Omega_0$ is the energy of the phonon involved in the formation of Holstein polarons, $g^2$ is related with EPC constant $\lambda$ by $g^2 = 2t\Omega_0\lambda$ and $t$ is the hopping amplitude.

In the spectral function simulation, a quadratic energy dependence of $\eta$ is adopted and a LL type spectral weight is included to mimic the experimental spectral function of *AB* band.

## Acknowledgment

The authors thank the helpful discussion with P. Z. Tang. This work was supported by the National Natural Science Foundation of China (Grants No. 11774190, No. 11427903 and No. 11634009), the National Key R&D program of China (Grants No. 2017YFA0304600, No. 2017YFA0305400, and No. 2017YFA0402900), and EPSRC Platform Grant (Grant No. EP/M020517/1). L. X. Y. acknowledges the support from Tsinghua University Initiative Scientific Research Program.

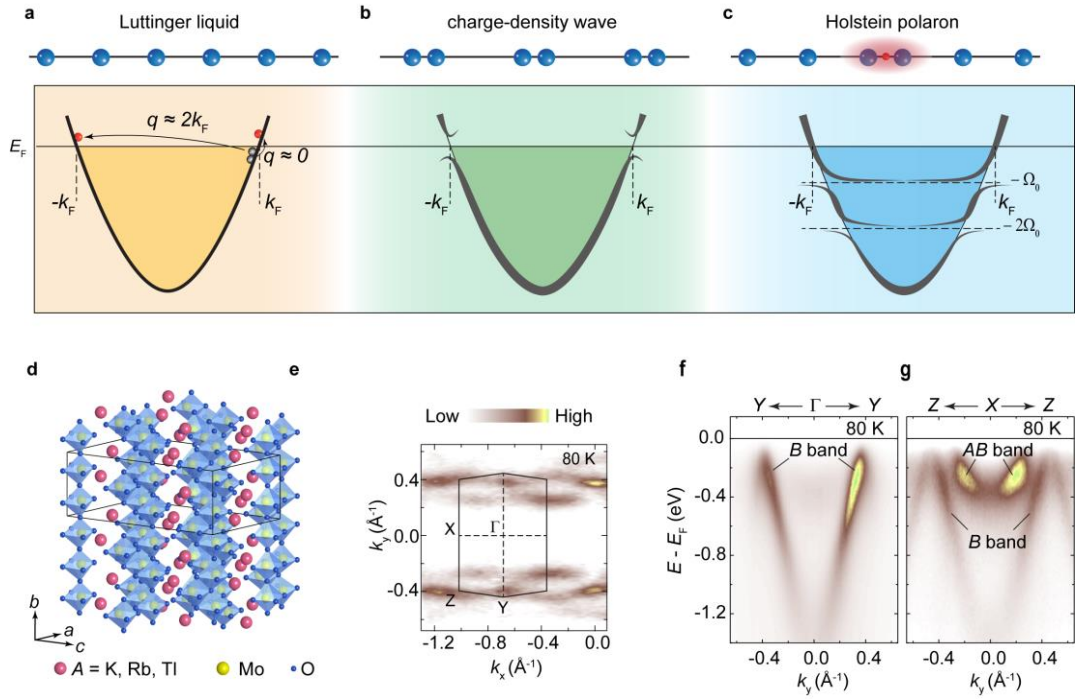

FIG.1. Physics in one dimension and basic band structure of $K_{0.3}MoO_3$. **a** Luttinger liquid behavior in one dimension, where only low-energy particle-hole pairs with wave vector $|q| \approx 0$ or $|q| \approx 2k_F$ can be excited. **b** Charge-density-wave (CDW) and periodic lattice distortion, leading to a gap opening near the Fermi energy ($E_F$). **c** An electron dressed by the local lattice distortion, forming the Holstein polaron and leading to band gaps and flat dispersions near multiples of the phonon energy ($\Omega_0$). **d** Crystal structure of $A_{0.3}MoO_3$ ($A$ = K, Rb, Tl) showing one-dimensional chains of $MoO_6$ octahedrons. Black lines indicate the unit cell. **e** Experimental Fermi surface (FS) by integrating ARPES intensity in an energy window of 20 meV near the Fermi energy ($E_F$), with surface Brillouin zone appended. **f, g** Band dispersions along $\Gamma$-$Y$ and $X$-$Z$ showing the bonding ($B$) and antibonding ($AB$) bands respectively. Data in **e**-**g** were collected at 80 K with a He lamp ($h\nu$=21.2 eV).

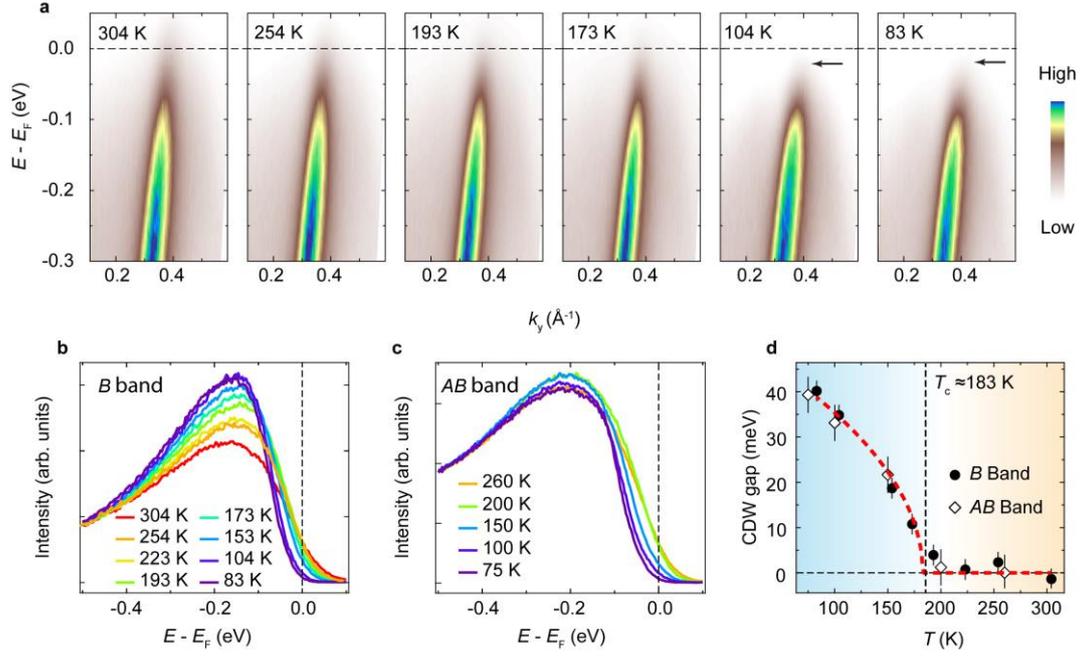

FIG.2. CDW transition in $K_{0.3}MoO_3$. **a** Temperature evolution of the band dispersion along the Γ-Y direction. The black arrows indicate the CDW gap opening below $T_{CDW}$ ≈ 183 K. **b**, **c** Energy distribution curves (EDCs) near Fermi momenta of B (**b**) and AB (**c**) bands at different temperatures showing the CDW gap opening below $T_{CDW}$. **d** CDW gap as a function of temperature. The CDW gap is determined by the shift of the EDC leading edge with respect to the spectrum at high temperatures. The red dashed line is the fit to the BCS-type gap equation. Data were collected with linear-horizontally polarized laser at 7 eV.

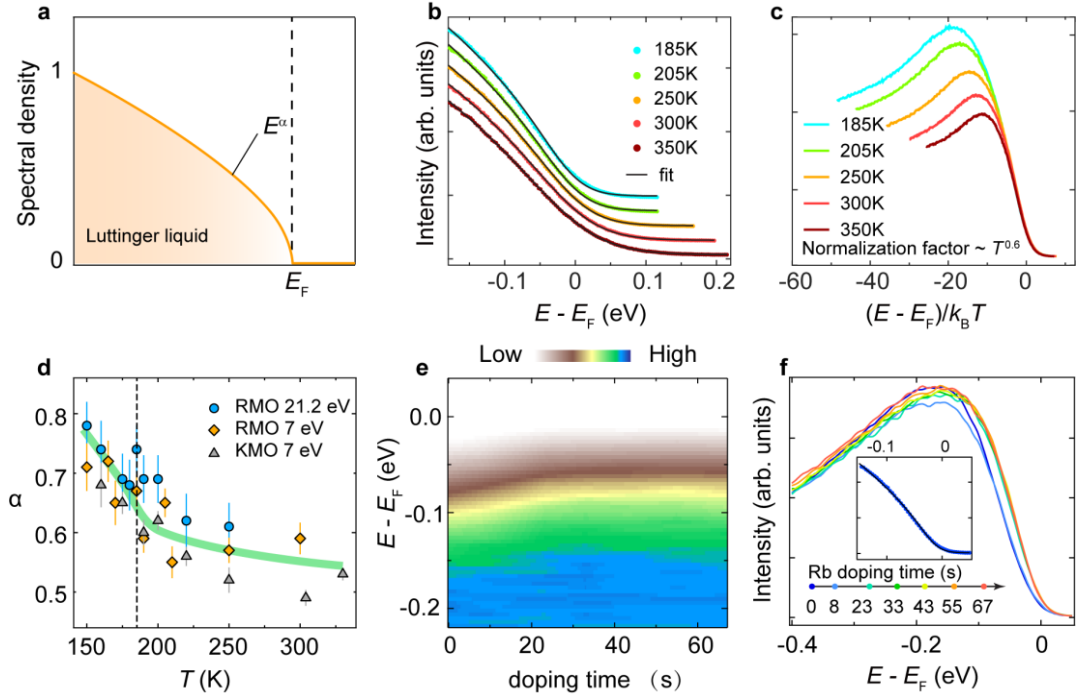

FIG.3. Luttinger liquid (LL) behavior of $A_{0.3}$MoO$_3$. **a** The spectral density of LLs (orange curve) showing a power-law dependence on the energy. **b** EDCs of Rb$_{0.3}$MoO$_3$ integrated along Γ-Y with fits to finite temperature LL model (black curves). The curves are vertically offset for clarity. **c** Scaling plot of the EDCs at different temperatures. The scaling factor is $T^\alpha$ with α = 0.6. **d** The anomalous exponent α as a function of temperature measured with He lamp (circles) and laser (diamonds and triangles). **e, f** Evolution of the EDCs near $k_F$ of the B band with surface Rb doping. The inset in **f** shows the fit of the integrated EDC after 6$^{th}$ (67 s) Rb doping to the LL model. Data in panels **b** and **c** were collected using 7 eV laser, while the data in panels **e** and **f** were collected using He lamp (Supplementary Fig. 6).

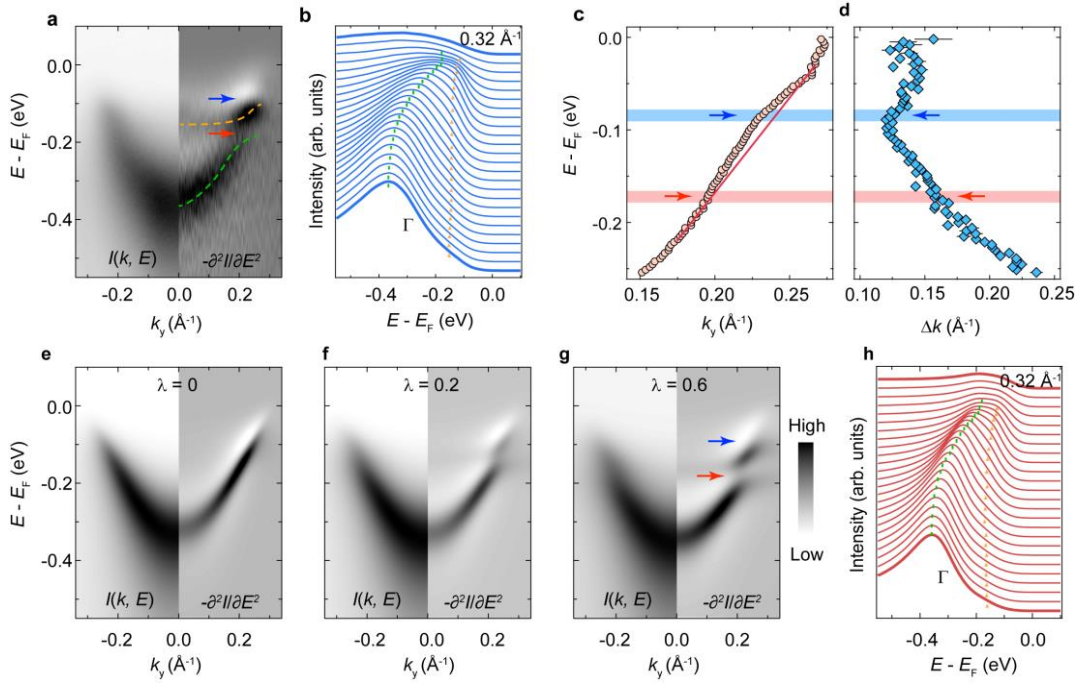

FIG.4. Holstein polarons in $K_{0.3}MoO_3$. **a** ARPES intensity map of the *AB* band along *X-Z* (left) and its second derivative (right) with respect to the energy axis measured with linear-vertically polarized 7 eV laser at 75 K. The orange and green dashed curves are the guides to eyes for the band discontinuity (emphasized with blue and red arrows). **b** Stacking plot of EDCs showing observed dispersions. The green bar and orange triangles are the guides to eyes for the dispersions. **c**, **d** Extracted *AB* band dispersion (**c**) and the spectral broadening (**d**) showing a kink-like feature near 85 meV below $E_F$ (blue arrow). The red line in panel **c** is the guide to eyes for the bare band dispersion. **e-g** Simulation of the spectral function (left) and its second derivative (right) of Holstein polarons at different EPC parameters. **h** Simulated EDCs showing good agreement with the raw data in **a**. The green bar and orange triangles are the guides to eyes for the dispersions.